\documentclass[aps,amsmath,amssymb,reprint,superscriptaddress,floatfix]{revtex4-2}

\usepackage{preamble}
\makeatletter
\DeclareRobustCommand{\equalcontrib}{%
  \frontmatter@footnote{These authors contributed equally.}%
  }
\makeatother

\begin{document}

\preprint{APS/123-QED}

\title{Few-photon degenerate parametric resonance in a two-tone driven microwave resonator}

\author{O.~Ameye\equalcontrib}
\email{orjan.ameye@uni-konstanz.de}
\affiliation{Department of Physics, University of Konstanz, 78464 Konstanz, Germany}

\author{J.D.~Koenig\equalcontrib}
\affiliation{Kavli Institute of Nanoscience, Delft University of Technology, PO Box 5046, 2600 GA Delft, The Netherlands}

\author{C.A.~Potts}
\affiliation{Department of Electrical and Software Engineering, University of Calgary, 2500 University Drive NW, Calgary, Alberta T2N 1N4, Canada}

\author{O.~Zilberberg}
\affiliation{Department of Physics, University of Konstanz, 78464 Konstanz, Germany}

\author{G.A.~Steele}
\email{g.a.steele@tudelft.nl}
\affiliation{Kavli Institute of Nanoscience, Delft University of Technology, PO Box 5046, 2600 GA Delft, The Netherlands}

\date{\today}

\begin{abstract}
Multi-tone external driving offers a route to parametric physics without directly modulating the device. However, the validity of the parametric response in the few-photon regime remains underexplored. Here, we apply two coherent microwave tones to a Josephson-junction Kerr oscillator and stimulate degenerate parametric downconversion via four-wave mixing. Using transmission spectroscopy, we observe that the response retains the qualitative semiclassical Kerr parametric oscillator structure, including its instability lobe and bistable phase-space topology. Interestingly, we demonstrate that a conventional single-mode reduction fails to capture the system quantitatively: the predicted AC Stark shift is severely underestimated, and the reported distributions might not be fully physical when the single-photon Kerr shift $K$ exceeds the cavity linewidth $\kappa$. Instead, we show that a full three-tone quantum description accurately reproduces the experimental observables. There, quantum fluctuations of the drive tones become dynamically dominant over dissipation, and all three interacting tones operate in a deep few-photon limit where the expected semiclassical macroscopic lobes undergo fundamental renormalization due to profound mixing with quantum variance. Our results establish two-tone-driven Kerr oscillators as potential parametric amplifiers and open new horizons to explore the quantum-to-classical crossover in driven-dissipative circuits.
\end{abstract}

\maketitle

Degenerate parametric driving constitutes a foundational mechanism across modern physics~\cite{Eichler2023Classical}. There, modulation of a device near twice its fundamental resonance stimulates coherent energy exchange and generation of degenerate excitation pairs. Consequently, these dynamics routinely supply squeezed states and instability lobes across diverse optical and mechanical platforms~\cite{Aasi2013Enhanced, Marhic2015Fiber, Pirkkalainen2015Squeezing,marti2024Quantum,Eichler2023Classical}. Among such platforms, superconducting Josephson circuits host Kerr parametric oscillators (KPOs) endowed with the strongest per-photon Kerr nonlinearity. Correspondingly, a handful of photons can already explore the strongly nonlinear physics of the system. Superconducting Josephson circuits have enabled the stabilization of Kerr-cat qubits~\cite{Grimm2020Stabilization, Frattini2017,Lescanne2020,Berdou2023,Reglade2024}, and they provide quantum-limited microwave amplification and single-shot qubit readout~\cite{Yurke1988,Castellanos2008, Aumentado2020Superconducting,Blais2021Circuit}. Additionally, KPOs facilitate the observation of driven-dissipative phase transitions~\cite{Fitzpatrick2017,benary2022experimental,Chen2023,beaulieu2025observation}. Alongside these experimental advances, we note a sustained theoretical mapping of their multistability, parametric instabilities, and squeezing landscapes~\cite{papariello2016ultrasensitive,Leuch2016Parametric,bartolo2016Exact,Heugel2019Classical, Wustmann2019, Minganti2018, Soriente2021,Ameye2025Parametric}. This robust framework directly supports broader programmes targeting bosonic-code qubits and reservoir-engineered nonclassical stationary states~\cite{Puri2017,Wang2019,Alvarez2024Biased, Margiani2025Boltzmann}.

\begin{figure}[tbp]
  \centering
  \includegraphics[width=\linewidth]{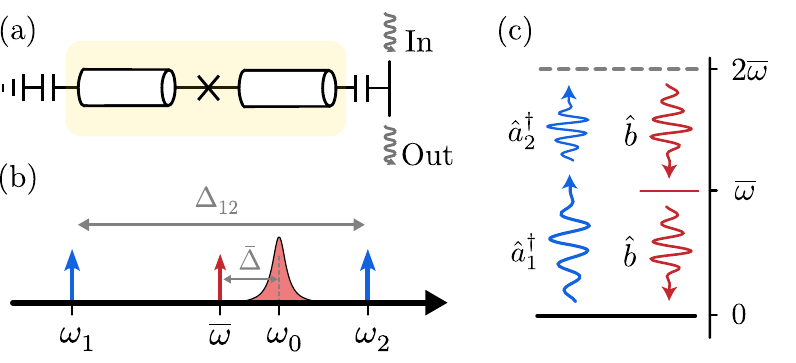}
  \caption{
    \textit{Device and drive scheme.} (a) Superconducting coplanar waveguide resonator with an embedded Josephson junction, acting as a nonlinear (Kerr) oscillator [cf.~Eq.~\eqref{eq: Kerr Hamiltonian}], addressed through the input/output port. (b) Frequency layout: two coherent microwave tones at $\omega_1$ and $\omega_2$ (blue arrows) are separated by $\Delta_{12}$ [cf.~Eq.~\eqref{eq: drive Hamiltonian}]; stimulated parametric drive at $\bar{\omega}$ (red arrow) manifests via four-wave mixing [see (c)] and is detuned from the bare oscillator resonance $\omega_0$ (red Lorenzian) by $\bar{\Delta}$. (c) Energy-level diagram of the underlying four-wave-mixing process: one photon is absorbed from each drive ($\hat{a}_1$, $\hat{a}_2$; blue) and two degenerate photons are emitted into the parametric tone $\hat{b}$ at frequency $\bar{\omega}$ (red).
  }
  \label{Fig:01}
\end{figure}

The dominant scheme for realizing the parametric drive in the few-photon Josephson literature is flux pumping a superconducting quantum interference device (SQUID). There, an external magnetic flux modulated at near twice the cavity frequency drives the parametric process through the flux-tunable inductance~\cite{Wustmann2019}. These schemes are powerful and well-developed. Crucially, however, they introduce severe hardware drawbacks: (i) they require dedicated on-chip flux lines, (ii) the resonator's Kerr coefficient and linewidth become functions of a DC flux bias that must be re-trimmed each cooldown~\cite{Krantz2019Guide}, and (iii) flux noise results in dephasing. At scale, these requirements impose critical thermodynamic and operational limitations. Specifically, each flux line adds to the cryogenic heat load and directly competes with the cooling budget~\cite{Hougland2025PumpEfficient,Dai2025PumpCoupling}. Furthermore, dense flux-biased arrays suffer from non-zero magnetic crosstalk between neighboring loops~\cite{Dai2021FluxCrosstalk}.

An alternative to flux modulation is multi-tone wave mixing~\cite{shen1983principles,Boyd2020}. This foundational mechanism of nonlinear optics combines multiple coherent drives in a nonlinear medium to generate new tones at their sum and difference frequencies. These interactions produce squeezing, entanglement, frequency conversion, and amplification across optics, microwaves, and mechanical systems~\cite{Eichler2023Classical}. Crucially, these processes are generated entirely through the dynamics of the drives in the medium's intrinsic nonlinearity, leaving the bare device untouched. Specifically relevant to our work, a two-tone drive can act as an effective degenerate parametric drive at the midpoint frequency via four-wave mixing~\cite{Kamal2009,boutin2017effect,yongxu2025Parametric}. However, it remains open whether such driving reduces to a single-mode KPO description in the few-photon quantum regime. Hence, a direct test is required to establish if this reduction remains quantitatively predictive. Such validation is essential for the operational regime of bosonic-code qubits and Kerr-cat protection.

In this work, we experimentally characterize stimulated degenerate parametric amplification in a two-tone-driven Josephson-junction Kerr oscillator in the few-photon limit. Using transmission spectroscopy, we directly test the validity of the conventional single-KPO reduction. We demonstrate that this standard framework fails to quantitatively capture the system. This failure stems from a strict physical boundary: the reduced model discards the drive-tones' quantum fluctuations, thereby underestimating the AC Stark shift and missing the saturation of the parametric population. Instead, using a full three-tone quantum description, we accurately reproduce the experimental observables. Crucially, all three interacting tones operate in a deep few-photon limit, i.e., the quantum variance of the drives becomes dynamically dominant over dissipation. Consequently, the expected semiclassical instability lobes undergo fundamental renormalization. Ultimately, our results establish two-tone driven Kerr oscillators as a robust platform to explore the quantum-to-classical crossover in driven-dissipative circuits.

\begin{figure}[tbp]
  \centering
  \includegraphics[width=\linewidth]{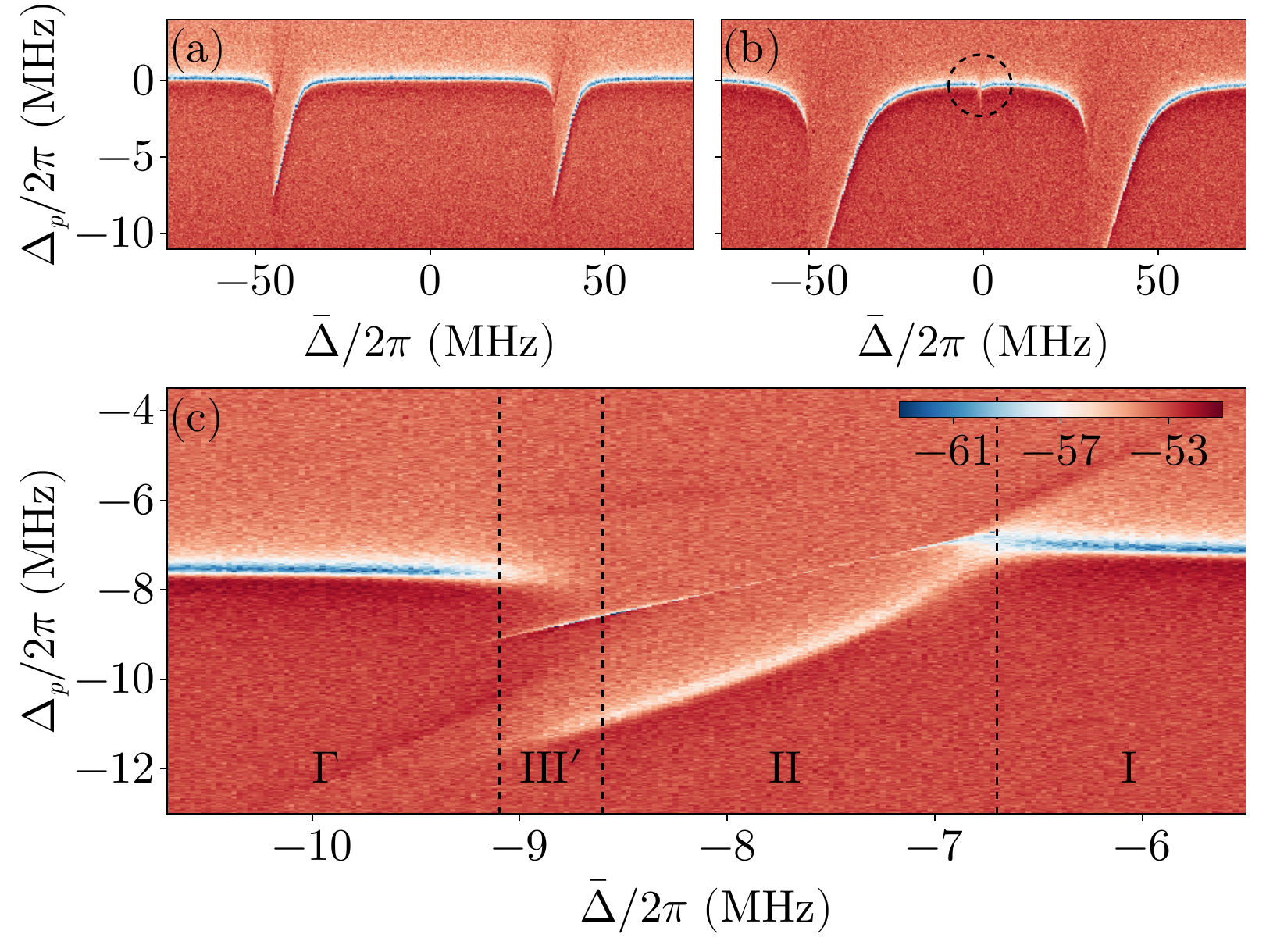}
  \caption{
    \textit{Degenerate parametric excitation in a two-tone driven Kerr nonlinear oscillator.} Measured transmission $|S_{21}|$ (color, dB) as a function of probe detuning $\Delta_p=\omega_{\rm pr}-\omega_0$ and drive midpoint detuning. (a) Transmission spectrum at low drive power ($-20.0$~dBm at the generator, tone separation $\Delta_{12}=80$~MHz). The probe resolves the bare cavity dip when the two-tone pump is off-resonance. As either pump tone crosses resonance, it drives a high-amplitude Duffing state. There, the linear response is renormalized via the AC Stark effect. This manifests as signal and idler side-peaks bending away from the bare resonance. (b) Spectrum at intermediate drive power ($-8.0$~dBm at the generator, $\Delta_{12}=80$~MHz). An additional signature emerges near the drive midpoint $\bar{\omega}$ (dashed circle). This indicates the onset of the degenerate parametric response. (c) High-resolution spectrum of the midpoint region at elevated drive power ($+12.0$~dBm at the generator, $\Delta_{12}=200$~MHz). The transmission displays distinct features delimiting the regions of the effective-KPO phase diagram (vertical dashed lines and labels), cf.~Fig.~\ref{Fig:03} and Eq.~\eqref{eq: effective KPO}. Region III${}^\prime$ denotes the sub-interval where the vacuum state contributes visibly alongside the bright parametric state.
  }
  \label{Fig:02}
\end{figure}

Our device is a superconducting coplanar waveguide resonator with an embedded Josephson junction, see Fig.~\ref{Fig:01}(a). We focus therein on a single microwave mode that realizes a nonlinear Duffing resonator,
\begin{align} \label{eq: Kerr Hamiltonian}
  \hat{H}_0/\, \hbar = - \Delta \hat{a}^\dagger \hat{a} + K \hat{a}^\dagger \hat{a}^\dagger \hat{a} \hat{a}\,,
\end{align}
where $\hat{a}$ annihilates an excitation at the bare resonance $\omega_0/2\pi = 6.5477$ GHz, $\Delta=\omega-\omega_0$ defines the detuning relative to a direct calibration drive~\cite{supmat}, and $K /2\pi\approx - 523$ kHz sets the Kerr nonlinearity~\cite{vool2017introduction}. The mode couples to a feedline with an external rate $\kappa_{\rm e} /2\pi\approx  275$ kHz alongside an intrinsic loss $\kappa_{\rm i} /2\pi \approx  49$ kHz, yielding a total energy decay $\kappa /2\pi \approx  324$ kHz. Cooled to $10$ mK, the system dynamics follow the master equation $\dot{\hat{\rho}} = -i[\hat{H}_0,\, \hat{\rho}]/\hbar + \kappa \mathcal{D}[\hat{a}]\hat{\rho}$~\cite{Carmichael1993,Breuer2007}.

In a side-coupled geometry, we drive the KPO with two coherent microwave tones at frequencies $\omega_1$ and $\omega_2$ separated by $\Delta_{12}$, see Fig.~\ref{Fig:01}(b). The drives in the laboratory frame read
\begin{align}
  \hat{H}_{\rm F}/\,\hbar = F (\hat{a}^\dagger + \hat{a}) \left[\cos(\omega_1 t) + \cos(\omega_2 t)\right]\,,
  \label{eq: drive Hamiltonian}
\end{align}
with shared amplitude $F$. These strong tones drive the mode into a non-equilibrium stationary state (NESS)~\cite{Ikeda2020General,Ikeda2021Nonequilibrium,Mori2023Floquet}. To probe this NESS, we superimpose a weak tone at $\omega_{\rm pr}$ and measure the complex transmission $S_{21}(\omega_{\rm pr})$~\cite{Gardiner1985,Lecocq2017Nonreciprocal, Blais2021Circuit,Arfini2025Magnon}. Away from frequencies coherently emitted by the pumped NESS, this coefficient maps directly to the dynamical susceptibility and captures the linear response in the laboratory frame rather than a rotating frame~\cite{Clerk2010,Heugel2023Role}. At the drive midpoint, however, the probe-induced midpoint field interferes coherently with the transmitted probe and modifies the measured transmission~\cite{supmat}.

In Fig.~\ref{Fig:02}, we show $\abs{S_{21}}$ against the probe detuning $\Delta_p=\omega_{\rm pr}-\omega_0$ and the drive midpoint detuning $\bar{\Delta}=(\omega_1+\omega_2)/2-\omega_0$. Unpumped, the probe resolves the bare cavity resonance at $\omega_0$. At low pump power, as either tone crosses resonance, it drives a high-amplitude NESS, see Fig.~\ref{Fig:02}(a). There, the AC Stark effect renormalizes the linear response, producing frequency-shifted signal and idler side-peaks that bend away from resonance~\cite{Huber2020Spectral,Heugel2023Role,Fu2025Sideband}. At intermediate power, an additional signature emerges near $\bar{\Delta}\sim 0$, see Fig.~\ref{Fig:02}(b). The Kerr nonlinearity mediates a four-wave mixing process, converting one photon from each drive into a degenerate pair in the midpoint tone~\cite{Kamal2009,boutin2017effect,yongxu2025Parametric}. This interaction implies a Hamiltonian term $\propto K \hat{a}_1 \hat{a}_2 \hat{b}^\dagger \hat{b}^\dagger + {\rm h.c.}$, where $\hat{b}$ annihilates a photon at $\bar{\omega}$, cf.~Fig.~\ref{Fig:01}(c).

We resolve the parametric response at high drive power, see Fig.~\ref{Fig:02}(c).  Signal and idler modes of a squeezed cavity vacuum state emerge around $\Delta_p\sim \bar{\Delta}$ in region I, and reappear in region $\Gamma$ with reversed spectral order. Decreasing the detuning from region I, the sidebands converge and merge into region II. There, the transmission indicates the onset of a high-amplitude state, displaying two symmetrically spaced side-peaks alongside a sharp central signature. As $\bar{\Delta}$ is swept lower, we enter region III${}^\prime$, where the features of region II coexist with the side-peaks of region $\Gamma$. We note that the system response now rests upon an enhanced AC Stark shift background, as evidenced by the horizontal baseline stabilizing near $\Delta_p/2\pi \sim -8$ MHz.

\begin{figure}[tbp]
  \centering
  \includegraphics[width=\linewidth]{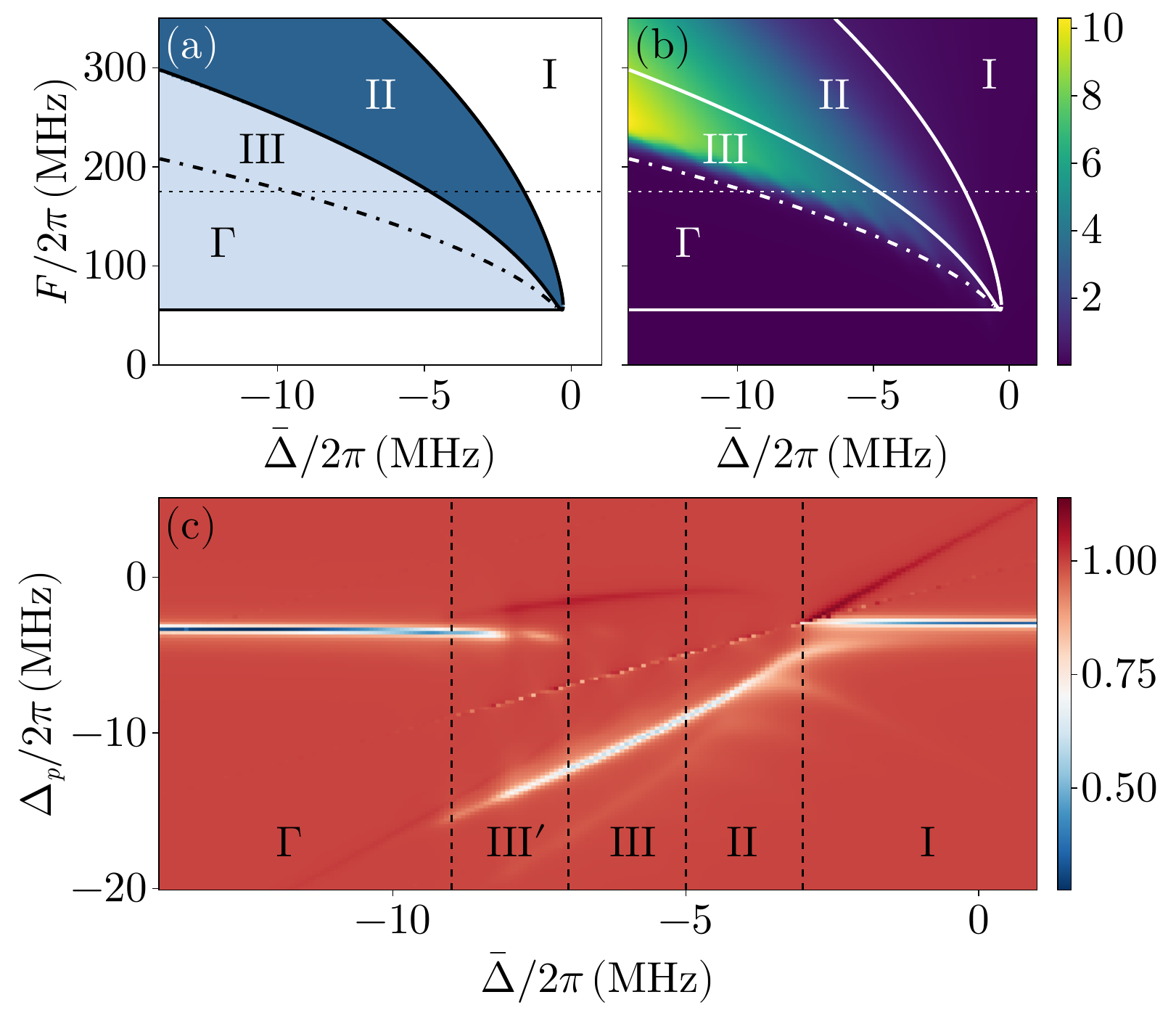}
  \caption{\textit{Harmonic balance results [cf.~Eqs.~\eqref{eq: three-tone model}-\eqref{eq: mixed transmission}].}  (a) Mean-field phase diagram as a function of two-tone drive strength $F$ and midpoint detuning $\bar{\Delta}$, where the order parameter, $\beta=\langle\hat{b}\rangle$, takes values of: vacuum only (white, region I), two spontaneous $\mathbb{Z}_2$ symmetry-broken bright parametric phase states (dark blue, region II), and coexistence of the vacuum with the two phase states (light blue, regions III and $\Gamma$). (b) Stationary-state photon number $\expval{\hat{b}^\dagger\hat{b}}$ (color) from numerical single-mode KPO Lindblad master equation. Superposed white curves represent the mean-field phase boundaries of (a); the dash-dotted line marks the analytical dissipative phase transition separating regions III and $\Gamma$~\cite{DykmanDissipative2015,sepulcre2025analytical}. (c) Simulated weak-probe transmission $|S_{21}|$ (color) of the effective KPO as $\bar{\Delta}$ is swept at fixed $F/2\pi = 175$ MHz (dashed horizontal gray line in (a), (b)); vertical dashed lines mark the regions, with III${}^\prime$ denoting the sub-interval of region III in which the vacuum state contributes visibly to the transmission.
  }
  \label{Fig:03}
\end{figure}

To model the response, we use a Harmonic Balance approach to project the driven-dissipative Duffing master equation onto the three dominant Fourier components~\cite{shirley1965solution,grifoni1998driven,Kosata2022HarmonicBalance,Floquet2025Expansion,Bestler2026Quantum,supmat}, and obtain an effective three-tone model
\begin{equation}
H_{\rm HB}=\sum_i H_i + H_{\rm{d}} + H_{\rm int}\,,
\label{eq: three-tone model}
\end{equation}
where $\Delta_{1,2}=\bar{\Delta}\mp\Delta_{12}/2$, $\Delta_3=\bar{\Delta}$, and $H_i/\hbar=-\Delta_i\hat{c}_i^\dagger\hat{c}_i+K\hat{c}_i^{\dagger 2}\hat{c}_i^2$ for $\hat{c}_1=\hat{a}_1$, $\hat{c}_2=\hat{a}_2$, and $\hat{c}_3=\hat{b}$. The drive term $H_{\rm{d}}=H_{\rm F}$ is given by Eq.~\eqref{eq: drive Hamiltonian}. The interaction reads
\begin{align}
  H_{\rm int}/\hbar ={}& 4 K \left[ \left(\hat{a}_1^\dagger\hat{a}_1 + \hat{a}_2^\dagger\hat{a}_2\right)\hat{b}^\dagger\hat{b} + \hat{a}_1^\dagger\hat{a}_1\hat{a}_2^\dagger\hat{a}_2 \right] \nonumber \\
  &+ 2 K \left[ \hat{a}_1^\dagger\hat{a}_2^\dagger\hat{b}\hat{b} + {\rm h.c.} \right]\,,
\end{align}
describing cross-Kerr shifts and four-wave-mixing pair conversion. Combined with the tone dissipators $\kappa \mathcal{D}[\hat{a}_j]$ and $\kappa \mathcal{D}[\hat{b}]$, this yields three coupled Kerr modes with time-independent quantized amplitudes~\cite{Kosata2022HarmonicBalance,supmat}.

Dominated by external drives, we assume the pump modes to occupy coherent states with definite amplitude $\alpha_j\equiv\langle \hat{a}_j \rangle$~\cite{nigg2012Blackbox,minev2021Energyparticipation,zhang2019Engineering,basilewitsch2022Engineering,boutin2017effect}. Displacing the field by this background yields an effective single-mode KPO for the midpoint tone~\cite{supmat},
\begin{align} \label{eq: effective KPO}
  H_{\rm KPO}/\, \hbar = - \Delta_{\rm eff} \hat{b}^\dagger \hat{b} + K \hat{b}^\dagger \hat{b}^\dagger \hat{b} \hat{b} + \frac{G_{\rm eff}}{2} (\hat{b}^2 + \hat{b}^{\dagger 2})\,,
\end{align}
where $\Delta_{\rm eff}=\bar{\Delta}-4K(\abs{\alpha_1}^2+\abs{\alpha_2}^2)$ and $G_{\rm eff}=4K\alpha_1\alpha_2$ represent the drive-induced AC Stark shift and effective two-photon pump. Crucially, this secular mean-field reduction pins the drive modes entirely and discards their quantum fluctuations. In the following, we show that retaining this quantum variance is essential for quantitative agreement with the experiment.

We first solve the effective KPO under the mean-field approximation to find stationary solutions $\beta=\langle \hat{b} \rangle$~\cite{supmat}. In Fig.~\ref{Fig:03}(a), we map the stable phase boundaries. Far from parametric resonance (region I), only the vacuum state ($\beta=0$) is stable. In region II, the vacuum undergoes a $\mathbb{Z}_2$ spontaneous symmetry breaking into bright parametric phase states ($\abs{\beta}>0$)~\cite{Eichler2023Classical}. In regions III and $\Gamma$, both vacuum and bright solutions coexist. This closely resembles a conventionally driven KPO~\cite{Eichler2023Classical,Heugel2019Classical,Heugel2022Ising,Ameye2025Parametric, Alvarez2024Biased,Leuch2016Parametric}. However, as $F$ increases, the parametric-instability lobe shifts toward negative $\bar{\Delta}$, tracking the AC Stark shift in $\Delta_{\rm eff}$.

We solve the single-mode Lindblad master equation for $\hat{b}$ to capture beyond-mean-field dynamics. To this end, we assume a strong Kerr nonlinearity $K$ to support Hilbert-space truncation~\cite{Mercurio2025quantumtoolboxjl}. The stationary-state photon population $\expval{\hat{b}^\dagger\hat{b}}$ successfully reproduces the mean-field boundaries, see Fig.~\ref{Fig:03}(b). Specifically, tuning $\bar{\Delta}$ into region II crosses the parametric-instability boundary, converging toward the bright states with $\expval{\hat{b}^\dagger\hat{b}}\neq 0$. Concurrently, we identify a qualitative difference between the formalisms: the macroscopic population collapses back to $\expval{\hat{b}^\dagger\hat{b}}\sim 0$ at the boundary between regions III and $\Gamma$~\cite{DykmanDissipative2015,bartolo2016Exact,heugelQuantum2019, roberts2020Drivendissipative, sepulcre2025analytical}. This precisely matches the analytical dissipative phase transition boundary evaluated at $(\Delta_{\rm eff},G_{\rm eff})$~\cite{sepulcre2025analytical,supmat}. Mechanistically, the stationary density matrix can be approximately decomposed into conditional states localized in the semiclassical phase-space basins, $\rho_{ss}\approx \sum_k p_k \rho_k$, where $p_k$ is the basin weight and $\beta_k=\operatorname{Tr}(\hat{b}\rho_k)$ is its local complex amplitude. In the semiclassical limit, these conditional states are centered near the corresponding stationary solutions, but they need not be exact coherent states. Hence, while parametric phase states remain semiclassically stable, strong Kerr interactions and quantum fluctuations make their stationary contribution negligible.

Moving on, we compute the linear response on top of the stationary single-mode Lindblad solution, see Fig.~\ref{Fig:03}(c) and cf.~Fig.~\ref{Fig:02}(c). We do so by numerically evolving the density matrix in the presence of the probe drive, evaluating the $S_{21}(t)$ expressions, and Fourier transforming the result. The numerical results show excellent agreement with the experiment (albeit at reduced detuning), allowing us to reveal the origin of the measured signatures: as the stationary state manifests as a statistical mixture localized around the semiclassical attractors, the probe interrogates this statistical ensemble, and we can express the total linear transmission as the weighted sum over the local response profiles~\cite{supmat},
\begin{align}
    S_{21}(\omega_{\rm pr}) \approx \sum_k p_k S_{21}^{(k)}(\omega_{\rm pr})\,.
    \label{eq: mixed transmission}
\end{align}

\begin{figure}[tbp]
  \centering
  \includegraphics[width=\linewidth]{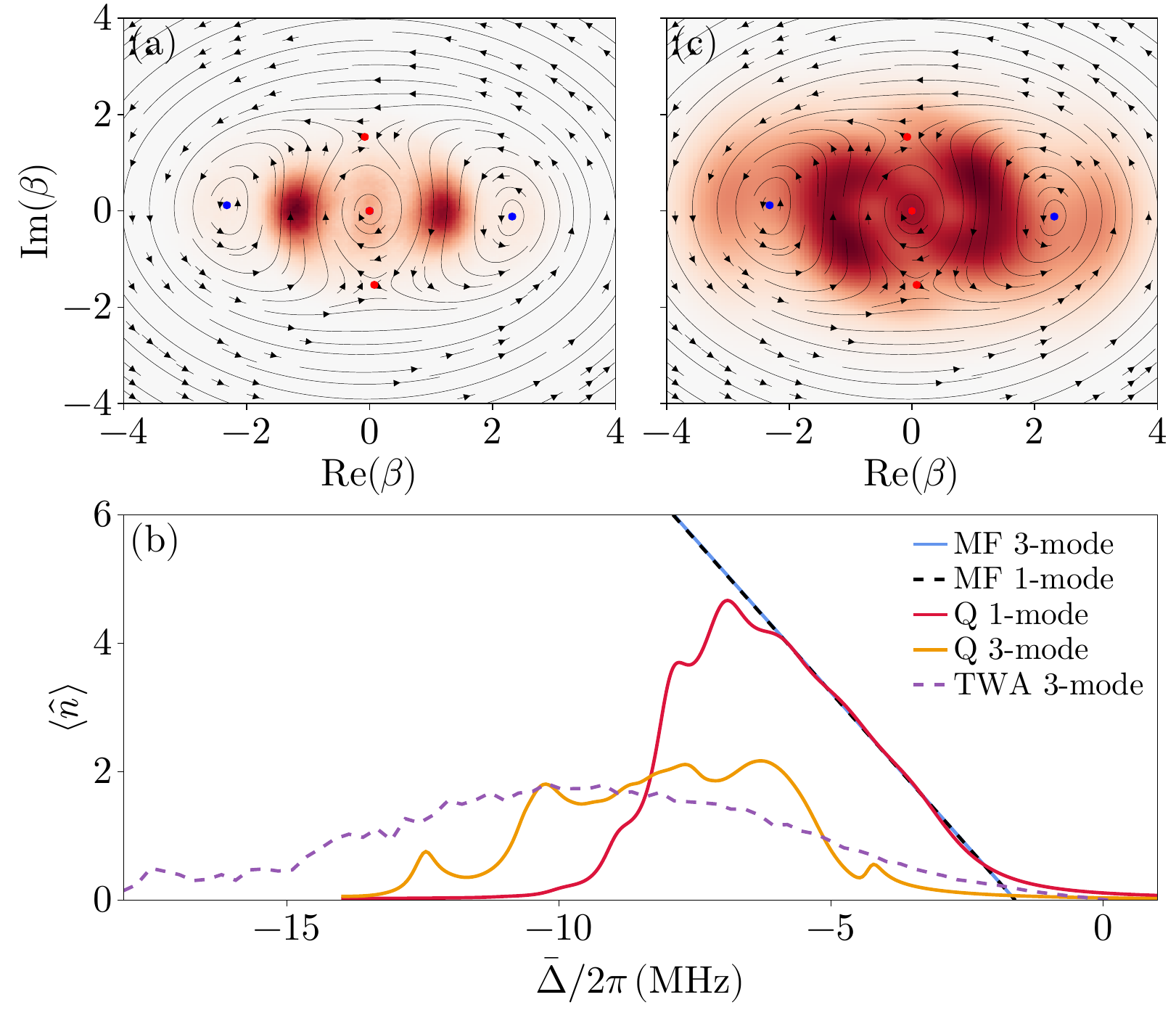}
  \caption{
    \textit{Model comparison of the parametric-mode occupation and phase-space distribution.} (a) Phase-space distribution of the parametric mode $\hat{b}$ at $\bar{\Delta}/2\pi=-7.26$ MHz computed from the three-mode TWA; blue dots mark the stable bright-state mean-field fixed points, red dots the corresponding unstable saddles, and streamlines the mean-field phase-space flow. The two Gaussian blobs characteristic of the $\mathbb{Z}_2$ parametric state sit between the origin and the bright-state fixed points, at lower amplitudes with respect to the mean-field prediction. (b) Stationary-state photon number $\expval{\hat{b}^\dagger\hat{b}}$ of the parametric mode versus $\bar{\Delta}$: mean-field (MF) and quantum Lindblad (Q) predictions for the single-mode KPO (1-mode) [cf.~Eq.~\eqref{eq: effective KPO}] and the three-mode harmonic-balance (3-mode) [cf.~Eq.~\eqref{eq: three-tone model}]  models, together with a Truncated Wigner Approximation (TWA) of the 3-mode model. (c) Wigner function obtained from the three-mode Lindblad master equation; the two Gaussians merge into an elongated bistable ridge.
  }
  \label{Fig:04}
\end{figure}

The local transmission profile $S_{21}^{(k)}$ captures two distinct scattering mechanisms. The standard cavity response generates distinct macroscopic sidebands at off-resonant probe frequencies, with weighted contributions from the vacuum and bright parametric states. Alongside this, the effective two-photon drive couples the signal and idler responses about both the vacuum and bright branches~\cite{Roy2016}. The bright parametric states additionally select one of the two symmetry-related amplitudes $\pm\beta$, spontaneously breaking the $\mathbb{Z}_2$ symmetry. When probed exactly at the drive midpoint ($\omega_{\rm pr}=\bar{\omega}$), the probe breaks the balance between the two symmetry-related bright states, producing a coherent midpoint response that occupies the same frequency channel as the transmitted probe and interferes with it to generate the central feature~\cite{supmat}. Away from the midpoint, this imbalance oscillates and is progressively suppressed because the state populations cannot follow it. The simultaneous stochastic sampling of the local responses manifests in region III$^\prime$, where the vacuum sidebands coexist with the interference and sidebands associated with the bright states. Moreover, the dissipative phase transition manifests as $p_{\rm bright}$ vanishes upon entering region $\Gamma$, just as quantum sampling suppresses $p_{\rm vacuum}$ within region III~\cite{bartolo2016Exact,heugelQuantum2019,seibold2026manifestations}.

The single-mode KPO successfully captures the qualitative spectral features. However, it substantially underestimates the pump-induced AC Stark shift. The effective detuning $\Delta_{\rm eff}$ predicts a $3$ MHz shift along the trajectory, whereas the measured transmission reveals a $6.8$ MHz shift, cf.~Fig.~\ref{Fig:02}(c). A corresponding discrepancy emerges in the output-field power spectral density~\cite{supmat}. These quantitative failures stem from the mean-field approximation: replacing each pump operator with its mean amplitude $\alpha_j$ entirely discards the quantum fluctuations. Mechanistically, the cross-Kerr interaction translates a single-photon fluctuation into a midpoint frequency shift of order $K$~\cite{Schuster2007Resolving,Gambetta2006Qubitphoton}. Consequently, the ratio $K/\kappa \approx 1.6$ in our device dictates that single-photon fluctuations exceed the cavity linewidth. This invalidates the mean-field reduction. Instead, we should expect the quantum variance of the pumps to dress the AC Stark shift, photon-number saturation, and phase-space distribution.

To model these discrepancies, we return to the full three-mode harmonic-balance model~\eqref{eq: three-tone model}. The mean-field reduction yields identical results to the single-mode KPO prediction, directly ruling out classical back-action. Instead, we apply the Truncated Wigner Approximation (TWA) to isolate the leading-order Gaussian pump fluctuations~\cite{polkovnikov2010Phase,steel1998Dynamical,sinatra2002Truncated,yoneya2025Pathintegral,mink2022Variational,supmat}. Essentially, we propagate the mean-field equations as stochastic differential equations subject to quantum fluctuation noise~\cite{rackauckas2017adaptive,DifferentialEquations.jl-2017}. The TWA reveals a modified phase-space population, see Figs.~\ref{Fig:04}(a) and (b). For example, in region II, the stationary TWA phase-space distribution exhibits the two Gaussian blobs characteristic of the parametric phase state~\cite{bartolo2016Exact,kheruntsyan1999Wigner,kryuchkyan1996Exact,heugelQuantum2019,Eichler2023Classical}, see Fig.~\ref{Fig:04}(a). However, quantum noise pulls the distributions inward toward the vacuum at $\beta=0$. This pump fluctuation renormalization manifests as a distinct saturation in the parametric photon number $\expval{\hat{b}^\dagger\hat{b}}$, see Fig.~\ref{Fig:04}(b). Crucially, the bright parametric window simultaneously shifts toward lower detunings, successfully recovering the enhanced AC Stark effect, cf.~Figs.~\ref{Fig:02}(c) and~\ref{Fig:04}(b).

As a next step, we solve the full three-mode model quantum mechanically via a Lindblad master equation. This successfully recovers both the additional AC Stark shift and the photon-number saturation, see Fig.~\ref{Fig:04}(b). Its qualitative agreement with the TWA supports the interpretation that pump-mode vacuum fluctuations drive these effects. In phase space, the full solution exhibits the same inward pull, while smearing the distinct bistable blobs into an elongated ridge, see Fig.~\ref{Fig:04}(c). Consequently, residual differences indicate that beyond-semiclassical quantum corrections remain relevant for detailed phase-space topology. Note that this full quantum treatment imposes a computational limitation: the density-matrix dimension scales as $n^{2\times 3}$ with the local Fock cutoff $n$, restricting simulations to small photon numbers and preventing quantitative parameter fits. Notwithstanding, within this truncated space, we conclusively demonstrate that the quantum variance of the pump occupations fundamentally dictates the experimental steady state.

In summary, we experimentally implemented a two-tone driven Kerr nonlinear oscillator to explore stimulated degenerate parametric amplification in the few-photon regime. We mapped its transmission spectrum using a hierarchy of theoretical models, identifying the mean-field phase diagram and demonstrating how a pump-induced AC Stark shift reshapes its instability lobe and sidebands. The qualitative KPO hallmarks survive at every theoretical level. Notwithstanding, our quantum simulations reveal a critical quantitative departure: the standard single-mode reduction completely fails to capture the effective parametric parameters. This highlights a fundamental deviation from semiclassical driving. Specifically, in the few-photon limit, the underlying four-wave-mixing process requires a quantum treatment that incorporates the applied pump fluctuations. Retaining this quantum variance correctly reproduces the enhanced Stark shifts, parametric photon-number saturation, and displaced bistable Wigner distributions. Consequently, our results establish two-tone driving as a robust tool to engineer and probe fluctuation-stabilized nonclassical stationary states.

\section*{Contributions}

J.D.K. performed the circuit theory work and simulations for the device design, produced the device design and fabricated the device, performed measurements, contributed to analysis theory and simulations, and contributed to writing of the manuscript. O.A. developed the theoretical framework and analytical derivations, performed all numerical simulations, helped interpret the experimental data, developed the physical interpretation that frames the work, and contributed to writing the manuscript. C.A.P. assisted in developing the theory and conceptual understanding, discussed intermediate results, supervised students, and helped prepare the manuscript. O.Z. supervised the theoretical work, co-developed the theoretical understanding and interpretation of the data, and took a leading role in developing the final form of the manuscript.  G.A.S. was involved in the conception of the experiment and the idea of using bichromatic driving, supervised the experiment, discussed intermediate results, contributed to developing the conceptual framework for understanding the data, helped develop the figures and the storyline, and gave feedback on the final manuscript. Data is available at \cite{zonedo}.

\vspace{2em}

\section*{Aknowledgments}

   We acknowledge the work of G.C.~Arends and S.~Leycuyer-Seguineau for initial contributions performing measurements characterizing devices. We thank J.~Maki and A.~Mikheev for valuable feedback and insightful discussions. O.A. and O.Z. acknowledge funding from the Deutsche Forschungsgemeinschaft (DFG) via project numbers 449653034 (Heisenberg), 425217212 (SFB1432), 521530974 (FOR5688), 545605411 (ANR), as well as from the Swiss National Science Foundation (SNSF) through the Sinergia Grant No.~CRSII5\_206008/1. J.D.K. and G.A.S. acknowledge financial support by the EU program H2020-FETOPEN project 828826 Quromorphic. C.A.P. acknowledges the support of the Natural Sciences and Engineering Research Council of Canada (NSERC) (No.~RGPIN-2026-05057).

\nocite{Dodonov2007,Everitt2009,Duffus2017,Berkeland1998Minimization}
\bibliography{bibliography}

\end{document}